\documentclass{article}
\usepackage[dvips]{graphicx}

\usepackage{latexsym}

\begin{document}

\pagestyle{plain} 
\setcounter{page}{1}
\setlength{\textheight}{700pt}
\setlength{\topmargin}{-40pt}
\setlength{\headheight}{0pt}
\setlength{\marginparwidth}{-10pt}
\setlength{\textwidth}{20cm}

\title{Effect of Closed Paths in Complex networks on Six Degrees of Separation and Disorder}
\author{Norihito Toyota   \and Hokkaido Information University, Ebetsu, Nisinopporo 59-2, Japan \and email :toyota@do-johodai.ac.jp }
\date{}
\maketitle

\begin{abstract}
Milgram Condition proposed by Aoyama et al. \cite{Aoyama} plays an important role on the analysis of "six degrees of separation". 
We have shown\cite{Toyota10}, \cite{Toyota11}  that the relations between Milgram condition and the generalized clustering coefficient, 
which was introduced  as an index for measuring the number of closed paths by us\cite{Toyota3}-\cite{Toyota7}, 
are absolutely different in scale free networks \cite{Albe2} and small world networks\cite{Watt1},\cite{Watt2}.
This fact implies that the effect of closed paths on information propagation is different in both networks. 
In this article, we first investigate the difference and pursuit what is a crucial mathematical quantity for information propagation.   
As a result we find that a sort of "disorder" plays more important role for information propagation  
than partially closed paths included in a network. 
Next we inquired into it in more detail by introducing two types of intermediate networks. 
Then we find that the average of the local clustering coefficient and the generalized clustering coefficients $C_{(q)}$
 have some different functions and important meanings, respectively. 
We also find that $C_{(q)}$ is close to the propagation of information on networks. 
 Lastly, we show that realizability of six degrees of separation in networks can be understood 
in a unified way by disorder.

 \end{abstract}
\begin{flushleft}
\textbf{keywords:}
 Milgram  Condition, Scale free network, Small world network, Generalized clustering coefficient, Entropy of degree 
\end{flushleft}
%\chapter{ss}
\section{Introduction}\label{intro}

\hspace{5mm} In 1967, Milgram has made a great impact on  the world by advocating "Six Degrees of Separation"\cite{Milg}. 
After that, his research group made some social experiments to establish the concept\cite{Milg2}, \cite{Milg3}.  
Recently Watts and his  research group  adduced evidence in support of the concept more intensively by experiments using e-mail\cite{Watt4}, \cite{Watt3}.

%%%%%%%%%%%%%%%%%%%%%%%%%%%%%%%%%%%%%%%%fffffffffffffffffffffffff
\begin{figure}[t]
%%%%%%%%%%%%%%%%%%%%%%% 
\begin{center}
\includegraphics[scale=0.9,clip]{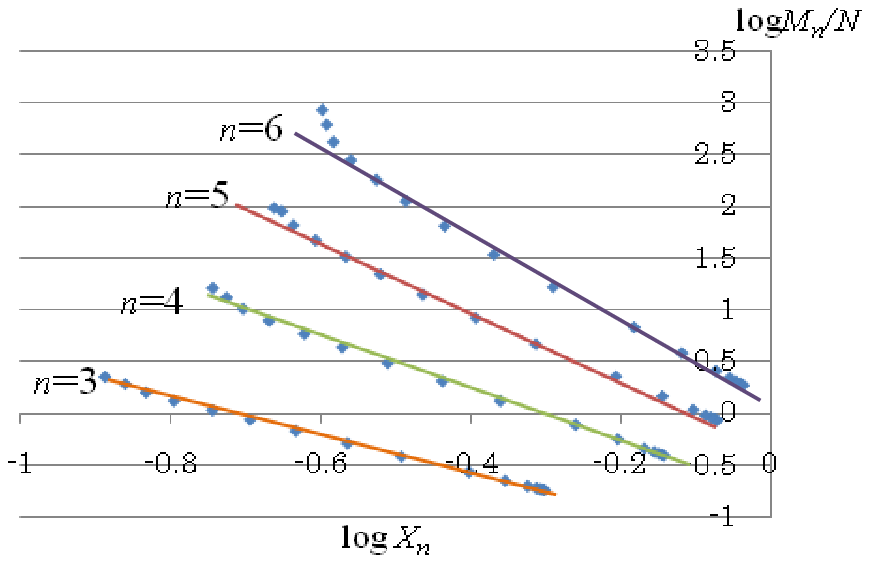} \\
%%%\end{center}
\caption{Sum of $C_{(p)}$ v.s.$ \log M_q/N$ in small world networks for every separation number $n$.}
\vspace{10mm}
%%%%%%%%%%%%%%%%%%%%
 %%\end{figure} 
%%\begin{figure}[b]
%%%%%%%%%%%%%%%%%%%%%%% 
%%%\begin{center}
\includegraphics[scale=0.9,clip]{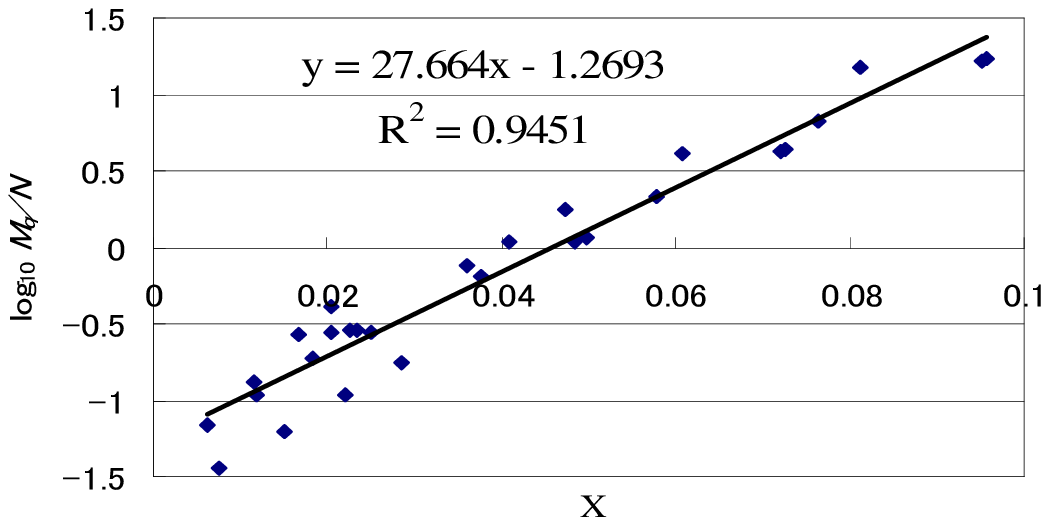}\\%%%%Fig6 in FIT 
\caption{ Separation number $q$ v.s.$ \log M_q/N$ in small-world networks with  several $\alpha$  }
\end{center}
%%%%%%%%%%%%%%%%%%%%
 \end{figure}

%%%%%%%%%%%%%%%%%%%%%%%%%%%%%%%%%%%%%%%%%%%%%%%%%%% 
%%%%%%%%%%%%%%%%%%%%%%%%%%%%%%%%%%%%%%%%fffffffffffffffffffffffff
%%%%%% \begin{figure}[t]
%%%%%%%%%%%%%%%%%%%%%%% 
%%%%%%\begin{center}
%%%%%%\includegraphics[scale=0.9,clip]{Fig5FITnew.eps} 
%%%\end{center}
%%%%%%\small
%%%%%%\caption{Sum of $C_{(q)}$ v.s.$ \log M_n/N$ in small world networks for every separation number $n$.}
%%%%%%\vspace{4mm}
%%%%%%%%%%%%%%%%%%%%
 %%\end{figure} 
%%\begin{figure}[b]
%%%%%%%%%%%%%%%%%%%%%%% 
%%%\begin{center}
%%%%%%\includegraphics[scale=0.7,clip]{SICEfig3.eps}%%%%Fig6 in FIT 
%%%%%%\tiny
%%%%%%\caption{ Sum of $C_{(q)}$ v.s.$ \log M_n/N$ in scale free networks.  }
%%%%%%\end{center}
%%%%%%%%%%%%%%%%%%%%
 %%%%%%\end{figure}
 %%%%%%%%%%%%%%%%%%%%%%%%%%%%%%%%%%%%%%%%
\normalsize 
We have studied the mathematical structures of complex networks \cite{Toyota3}-\cite{Toyota7} to understand "six degrees of separation" 
advocated by Milgram\cite{Milg}. 
 Especially, we have developed an original formulation \cite{Toyota3}-\cite{Toyota7} based on " string formulation" 
developed by Aoyama\cite{Aoyama} in order to study the effect of closed paths included in a network on information propagation.   
 
 We have proceeded with our study for scale free networks and small world networks in the basis of Milgram condition 
proposed by Aoyama \cite{Aoyama}. 
As a result, it was proved that the generalized clustering coefficient, which takes on the responsibility of closed cycles in a network, 
 has the opposite effect on information propagation in a sense of Milgram's experiment 
(The term of "information propagation" is always used  thoroughly in this sense in this article) in the both networks. 
That is, the existence of closed paths in a network impedes the propagation of information in small world networks, but 
promotes it in scale free networks. 
    
In this article, we first investigate the difference in both the networks and pursuit what is a crucial mathematical quantity for information propagation.   
As a result, we find that a sort of "disorder" plays more important role for information propagation rather 
than the generalized clustering coefficient and the average of local clustering coefficients.    
In fact, we show that an index responsible for information propagation is the entropy of the degree distribution and/or 
the standard deviation of degree data. 
Thus the opposite characteristic described above in the both networks can be understood in a unified way.

Next we further inquired into the subject in more detail by introducing two types of intermediate networks. 
One is the networks that link a regular network and a scale free network and second is the networks 
that link a sort of small world networks and the scale free network,  
It would be expected that the transition between  scale free network like property and small world network like one may be 
observed by such considerations.  

Then we find that the average of the local clustering coefficients $C^{(3)}$ and the generalized clustering coefficients $C_{(q)}$, 
which are proposed by us\cite{Toyota3}-\cite{Toyota7}  and have a generalization of the transitivity\cite{NewmB},\cite{Newm23},  
 have some different functions and important meanings, respectively. 
There is a positive or a negative correlation between $C_{(q)}$ and $M_6$ in Milgram condition which will be defined in the next section, that is, 
$C_{(q)}$ and information propagation on the whole,  
but $C^{(3)}$ may grasp the transition between scale free $\leftrightarrow$ small world (SF-SW transition). 
 Then we find that small entropy of degree distribution and small standard deviation of degree data impede the propagation 
of information all in small world networks, scale free networks and  the models introduced in this article. 
 So the behaviors of $M_6$ or six degrees of separation in all networks discussed in this article are 
understood by disorder.  

This article is planned as follows. 
After the introduction of the section 1, we describe the details of SF-SW transition in the section 2 
and discuss disorder of networks by introducing the entropy of a degree distribution and the standard deviation of degree data.
In section 3, we introduce two types of intermediate networks and estimate some statistical and global indices of the networks 
such as the average of local clustering coefficients, the generalized clustering coefficient, the average shortest path length 
and a quantity related to six degrees of separation by computer simulations.  
Moreover we estimate the degree of disorder of the two types of networks.  
In consequence, we show that the behaviors of $M_6$ or six degrees of separation in all networks discussed in this article are 
understood in a unified way by disorder.

\section{Scale free networks vs. Small world networks} 
\subsection{Milgram Conditionin Scale free networks vs. Small world networks}
\hspace{5mm}
According to Aoyama\cite{Aoyama}, we call $j$-$string$ string like part consisting of $j$ nodes in a network and 
$S_j$ is the number of $j$-$string$ in the network.  
 $\bar{S}_j$ is the number of non-degenerate $j$-$string$s \cite{Toyota3}-\cite{Toyota5}, which are  
closed strings or open strings.  
By using it, We have defined the generalized clustering coefficient $C_{(q)}$\cite{Toyota3}-\cite{Toyota5} as 
\begin{equation}
C_{(q)}=\frac{2q\times \;number \;of \;polygons }{number \;of\; connected \;q\mbox{-}plets }=\frac{ 2q\Delta_q  }{\bar{S}_q}, 
\end{equation}
where $\Delta_q$ is the number of polygonal structures with $q$ edges in a network. 
This definition is the same as usual global clustering coefficient or "transitivity"\cite{Wass} at $q=3$ and gives its extensions for $q>3$.  
This is different from the original clustering coefficient $C^{(3)}$ introduced by Watts\cite{Watt1},\cite{Watt2}, 
which is the average of the local clustering coefficients; 
\begin{eqnarray}
C^{(3)}_i &=& \frac{number \;of \;pairs  \;of  \;neighbors \; of \;node \; i \; that  \;connected}{number \;of \;pairs \;of \;neighbors \; of \;node \; i},  \nonumber \\ 
C^{(3)} &=& \frac{1}{N} \sum_i^N  C^{(3)}_i, 
 \end{eqnarray}
where $N$ is the network size which is defined by the number of nodes in a network.  
A series of $C_{(q)}$ can be described by the adjacency matrix in a systematic way \cite{Toyota3}-\cite{Toyota5}. 
Then Milgram condition for $n$-th degrees of separation in a network with the size $N$  is given by the  following formula \cite{Aoyama};
\begin{equation}
M_{n} \equiv \frac{\bar{S}_{n}}{N} \sim O(N).
\end{equation}    
When this condition is satisfied, it means that $n$-th degrees of separation  becomes possible in the network. 
As the left hand side of eq. (3) becomes larger, it becomes  easier to realize $n$-th degrees of separation. 

In the fig.1 and fig.2, the relations between $M_n/N$ and 
\begin{eqnarray}
X_n &\equiv& \sum_{p=3}^{n} C_{(q)} 
\end{eqnarray}
are shown for small world networks constructed by Newman-Watts model \cite{Newm22},\cite{Mona} 
and scale free networks constructed by the configuration like model where self-edges and multiedges are excluded in 
 the original configuration model\cite{Newm23} .

Fig.1 and fig.2 show that as the generalized clustering coefficient $C_{(q)}$ grows larger, 
it becomes easy for information to propagate in scale free networks, while it is difficult for information 
to propagate for larger$C_{(q)}$ in small world networks, provided, however, that data points with large $M_6/N$ are those at small scaling exponent $\alpha$ and
on the contrary the data points  with small $M_6/N$ are those at large $\alpha$ in fig.2. 
The same behaviors also apply for $C^{(3)}$. 

The aspect of them is quantitatively shown as
\begin{eqnarray}
M_n &\sim & B_n (X_n)^{-a},  \;\;\;   \mbox{for small world networks}\\
M_n &\sim & \exp( bX_n),   \;\;\;\;\;  \mbox{for scale free  networks}
\end{eqnarray}
where $B_n$, $a>0$ and $b>0$ are some constants\cite{Toyota10}, \cite{Toyota11}. 
$C_{(q)}$ depends on the scaling exponent $\alpha$ in scale free networks and 
as $\alpha$ becomes smaller\cite{Toyota10},  $C_{(q)}$ becomes larger, which means that it becomes easier for information to propagate.     
This fact is also suggested in the configuration model where the average path length $L$ for scale free networks with $N$ nodes is given\cite{Cohen}  by
\begin{eqnarray}
L\;\;\; \sim& \log \log N, \;\;\;\;\;\;\; & \mbox{for } 2<\alpha <3, \nonumber \\
  \sim& \log N /\log \log N, \;\;\;\;\;\;\;&  \mbox{for } \alpha =3, \nonumber \\
  \sim & \log N, \;\;\;\;\;\;\; &  \mbox{for } \alpha >3, 
\end{eqnarray}
and the average of the local clustering coefficients $C^{(3)}$ is given \cite{NewmB} by 
\begin{equation}
C^{(3)}=\frac{(<k^2>-<k>)^2}{<k>^3N}, 
\end{equation} 
where $<k>$ is the average degree and so on. 
Eq.(7) means that the smaller $\alpha$ is, the smaller $L$ is at a fixed $N$, so that 
information is considered to be easy to propagate over a network at smaller $\alpha$.  
As for $C^{(3)}$, it is suggested to be large at small $\alpha$ from eq. (8), because 
$<k^2>$ notably becomes larger than $<k>$ at small $\alpha$ by the cut-off theory\cite{Math} for which a brief review is given in the appendix. 
This consideration is consistent with the results of  the approximate configuration model used in our analysis. 

As said above,  the effect of the generalized clustering coefficient, which bears a responsibility on how many closed paths there is in a network, 
on propagation of information is just opposite in the scale free networks and the small-world networks.     
How can we understand this some strange phenomenon at a glance?
This will be discussed in the next subsection. 

\subsection{Disorder}
\hspace{5mm}
We focus on when $M_n$ is large in the both networks. 
$M_n$ becomes larger at large rewiring ratio $q$ in small world networks and then the networks come  closer to random networks. 
Increasing  the rewiring ratio makes  information propagation urge in small world networks. 
As for scale free networks constructed by the configuration model, $M_n$ becomes larger at small scaling exponent $\alpha$ and 
so decreasing $\alpha$ makes information propagation urge. 
Then $\alpha$ decreases,  it  also makes networks becomes disordered or homogeneous in degree.  
Thus we conjecture that the crucial point for information propagation would be a sort of "disorder".  
In both cases, "disorder", which is not yet defined appropriately, increase at least, as $M_n$ becomes larger.   
We consider two indices for studying information propagation on networks, the standard deviation $\sigma$ and 
the entropy as candidates of an amount of disorder. 

The standard deviation $\sigma$ of degree data is defined by 
\begin{equation}
\sigma^2=<k^2>-<k>^2.  
\end{equation} 

The entropy is known as a concept that represents a sort of disorder for physics 
or indefiniteness for information theory. 
Some similar concepts of entropy regarding networks have been proposed\cite{Sole},\cite{Bianc},\cite{LIJi}.  
The entropies introduced by \cite{Wang} and \cite{Sole} among them have analogous properties. 
In this article, we consider the entropy of degree distribution introduced by \cite{Wang} as an index relevant to disorder of networks;
\begin{equation}
S=\sum_{k=1}^{k_{max}}P(k)\log_2P(k), 
\end{equation}
where $P(k)$ is the degree distribution.  

For scale free networks, the entropy $S$ is estimated as the following;  
\begin{equation}
S=-\int _{k=k_{min}}^{\infty} dk P(k) \log_2 P(k)
 =\int_{k=k_{min}}^{\infty} dk  ck^{-\alpha} \log_2 (ck^{-\alpha} )  
 = -\Bigl( \log_2\bigl( \frac{\alpha-1}{k_{min}} \bigr)+\frac{\alpha}{1-\alpha} \Bigr),  
\end{equation}
where $\alpha>1$ and eq.(18) given in the appendix is used in the last equality. 
A typical data driven from eq.(15) in the appendix at $k_{min}=4$ are plotted in fig.3.  
As is shown in this figure, when $\alpha$ increases,  the entropy $S$ decreases. 

 \begin{figure}[t]
%%%%%%%%%%%%%%%%%%%%%%% 
\begin{center}
\includegraphics[scale=0.7,clip]{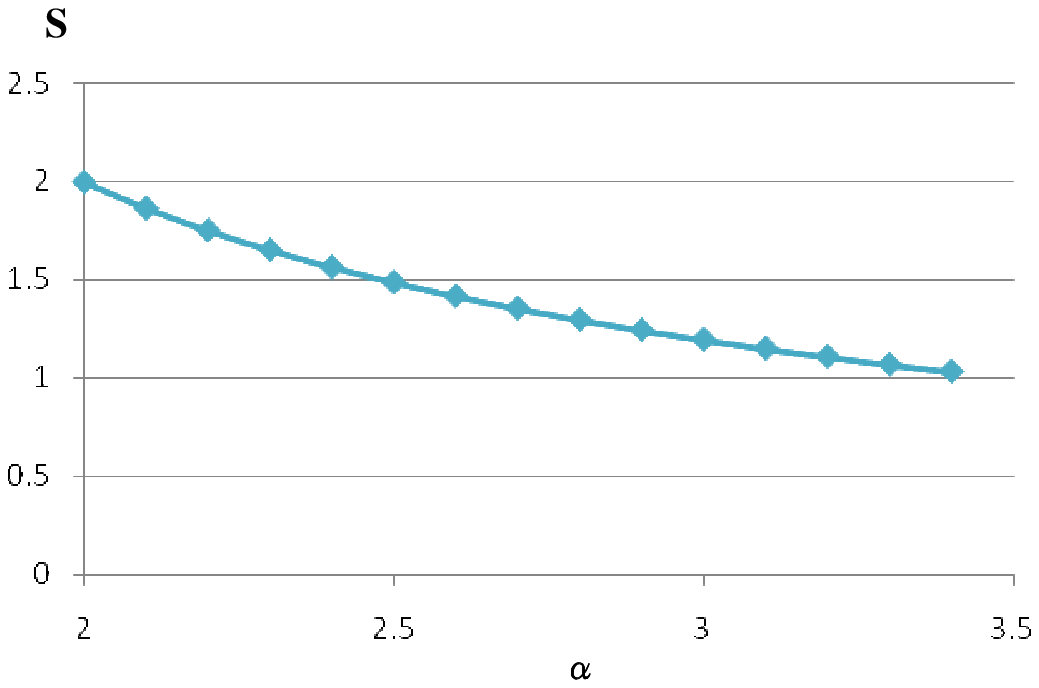} \\
{Entropy vs. scaling exponent in scale free networks for $k_{min}=4$.  }
\end{center}
%%%%%%%%%%%%%%%%%%%%
\end{figure} 

$S$ has the minimum value in small world networks at $q=0$, because all nodes have the same degree at $q=0$, that is, 
the network becomes a regular circle lattice (WS lattice). 
It is also well known that an entropy has larger value for random data.   
The larger $q$ is, the larger $S$ is.    
$\sigma$ also takes the minimum value $0$ at $q=0$ in small-world networks, because all nodes have 
the same degree at $q=0$.  
$q\neq0$ leads to large deviation in degree so that $\sigma$ has non-zero value.   

For scale free networks, we get from eq.(32)-(36) in the appendix, 
\begin{eqnarray}
<k>^2 \sim& \bigl( \frac{\alpha-1}{2-\alpha}\bigr)^2 k_{min}^2  N^{\frac{4-2\alpha}{\alpha-1}}, & \mbox{ for } 1<\alpha<2, \nonumber\\
\sim& \bigl( \frac{\alpha-1}{2-\alpha}\bigr)^2 k_{min}^2,  \;\;\;\;\; \;\;\;\;\;  & \mbox{ for } \alpha>2,
\end{eqnarray}
and
\begin{eqnarray}
<k^2> \sim &\frac{\alpha-1}{3-\alpha}  k_{min}^2 N^{\frac{3-\alpha}{\alpha-1}}, \;\;\;\;\; 
        & \mbox{ for } 1<\alpha<2, \nonumber\\
% c N^{3-\alpha}, \;\;\;\;\; & \mbox{ for } 1<\alpha<2, \nonumber\\
\sim &\frac{\alpha-1}{3-\alpha}  k_{min}^2,\;\;\;\;\; \;\;\;\;\; \;\;\;\;\; 
%cN^{(3-\alpha)/(\alpha-1)},  \;\;\;\;\; 
      & \mbox{ for } 2<\alpha<3. %\nonumber\\
      % &=const.  \;\;\;\;\; & \mbox{ for } 3<\alpha,
\end{eqnarray}

We can get a numerical result for $\sigma^2$ based on eq. (9). 
Fig.4 shows that $\sigma$ takes larger value at lower $\alpha$ except for near a singular point.   
The small peak at $\alpha \sim 3$ in fig.4 is due to the singularity inhered in $<k^2>$ shown in eq.(13).  
Such a singularity is idealized mathematical object and does not appear in real networks. 
  
Evaluated characters up to now in both networks are summarized in Table 1 
where $C$ originally shows $C_{(3)}$ but   $C^{(3)}$ also shows similar behavior. 
There are no quantities that correlate with $C$ but $\sigma$ and $S$ behave like 
$M_n$ qualitatively in both the networks.   
So it is noticed that it is $\sigma$ and $S$ that are relevant to information spreading, that is, Milgram condition. 
As the standard deviation and the entropy of degree distribution become larger, $M_n$ becomes larger, too. 
 Thus it is disorder relevant to information spreading 
in small world networks and scale free networks in common. 
As a consequence, we find a primary common factor in both networks.      
 
\begin{table}
\begin{center}
\begin{tabular}{c||c|c|c|c|c|c} \hline
Network &  $\alpha$ & $q$ & $M_n$ & $C$ & $S$ & $\sigma$ \\\hline
Small World &  & large & large  & small & large & large \\\hline
Scale Free & small &   & large & large & large & large \\\hline
\end{tabular}
\caption[Smsll World vs. Scale Free networks]{Small World vs. Scale Free networks}
\label{label}
\end{center}
\end{table}

\begin{figure}[t]
%%%%%%%%%%%%%%%%%%%%%%% 
\begin{center}
\includegraphics[scale=0.7,clip]{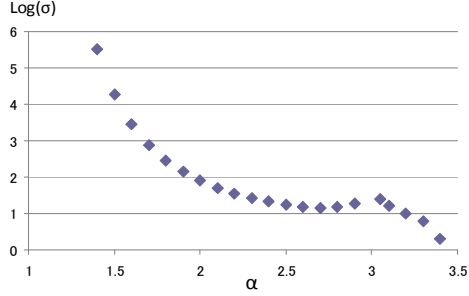} \\
\caption{Standard deviation in scale free networks for $N=400$.  }
\end{center}
%%%%%%%%%%%%%%%%%%%%
\end{figure}

\section{Other networks inspired by SF and SW networks}
\subsection{Two intermediate networks}

%%%%%%%%%%%%%%%%%%%%%%%
\begin{figure}[t]
\begin{center}
\includegraphics[scale=0.5,clip]{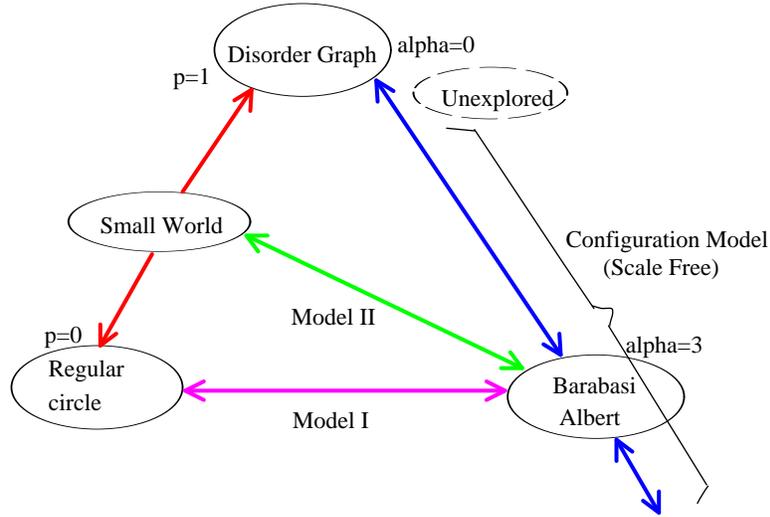} \\
\caption{The relation among networks inspired by scale free and small world networks.  }
\end{center}
\end{figure} 
%%%%%%%%%%%%%%%%%%%%
\hspace{5mm}
In order to investigate further the phenomenon that $C_{(q)}$ has  the opposite  
effect on information propagation in both networks, we introduce the following two types of intermediate networks linking scale free networks 
and small world or regular networks.  
A small world network is an intermediate network between a random network and a regular network in some way, which are controlled by a rewiring parameter $q$. 
The random network can be a disordered network in the meaning that its entropy is large.   
 On the contrary, the configuration models are  interpreted as intermediate networks between a disordered network that has a large entropy and 
Barabashi-Albert model\cite{Albe2} in some way, which are controlled by a scaling exponent $\alpha$.
In this article,  the term of "Barabashi-Albert model (BA model)", carries the connotation of scale free networks with scaling exponent $\alpha=3$, 
even though they are not constructed by using the preferential attachment.  
 Provided that scale free networks with $\alpha$ near zero can not studied practically because of bad singularity in $<k>$ and so on.  

So random networks are considered to be changed in two ways by parameter $q$ and $\alpha$, with the result that two types of changed networks 
(scale free networks and small world networks) have the opposite behavior in the effect of closed paths on  information spreading.  
The outline is described in fig. 5. 

We are motivated  to explore further two types of networks from the figure 5. 
They are called  Model I and II, that are intermediate networks between small world networks and BA model and between regular networks 
and  BA model, as described in fig.5, respectively.  
By investigating their intermediate networks, the critical phenomenon may be observed relevant to opposite behavior of $C_{(q)}$ in 
small-world networks and scale free networks. 
The constructions of these two types of networks are as follows. 

\begin{enumerate}
\item regular networks $\rightarrow$ scale free networks (Mode I)\\
  Preferentially attach $N-N_0$ nodes with the degree $k$ to nodes on a circle lattice (WS-lattice) 
with $N_0$ nodes which has all degree $k_0$.  $N-N_0$ is a control parameter in this model. 

\item Scale free networks $\rightarrow$ small world networks (Model II)\\
  Attach an edge  between two  nodes that connects to a hub,   
and instead deletes an edge attached with the hub but except for the edge between the hub and two nodes connected at present,
to conserve the total number of edges in the considering network. 
This procedure are done with probability $P$ for each node, which means that $PN$ nodes follow this procedure in the order of degree rank,  
and makes the clustering coefficient large.      
The rewiring ratio $P$ is introduced as a control parameter for clustering. 
There are a few attempts for constructing tunable scale free networks\cite{Wang},\cite{Holme},\cite{Zhang} that have small average path length 
and relatively large clustering coefficients. 
We do not accept them because they basically have a power low in the degree distribution that does not reflect 
the property of small world network. 
 The procedure in the model II, however, really corresponds to the third step in the model introduced by 
B.Wang et al. as referred later on\cite{Wang}. 
 \end{enumerate}

Since small world networks are a sort of the intermediate networks between a regular network and a random network,  
 it will be of interest in intermediate networks between scale free networks and regular networks, instead of random  networks.  
 It gives a motivation for studying Model I. 
 It will be also of interest to see what phenomena appear in the middle between scale free networks and small world networks 
with a large clustering coefficient, since $C_{(q)}$ has  the opposite effect on information propagation in both networks. 
It is the reason to investigate Model II.   

\subsection{Simulation Results of Model I}
\hspace{5mm}
(A), (B), (C) and (D) of fig. 6 show the average of local clustering coefficient $C^{(3)}$, $L$, sixth Milgram condition $M_6/N$ in Milgram condition and  
the general clustering coefficients $C_{(q)}$ for $3\leq n \leq 6$ vs. the initial size  $N_0$ of WS lattice in the model I,  respectively. 
The each plot is the average over ten rounds at $N=400$ for (A) and (B) or over five rounds at $N=200$ for (C) and (D) with $k=k_0=4$.   
The general clustering coefficients $C_{(q)}$ in fig.6 (D) are normalized as to be 1 at $N=200$ .  
 As it has been  shown that 
the behavior $M_6/N$ is well correlated with that of the average path length $L$ \cite{Toyota10},\cite{Toyota11},   
we also estimated $L$, which has an advantage over $M_n/N$ from the perspective of computational complexity, 
besides $M_n$.   
The reason that $M_6/N$ and $C_{(q)}$  are evaluated together over five rounds at $N=200$ is that almost same computational complexity is needed
 in calculating them.  
We, however, consider that the stable results  are given even in these cases.   

\begin{figure}[t]
%%%%%%%%%%%%%%%%%%%%%%% 
\begin{center}
\includegraphics[scale=0.75,clip]{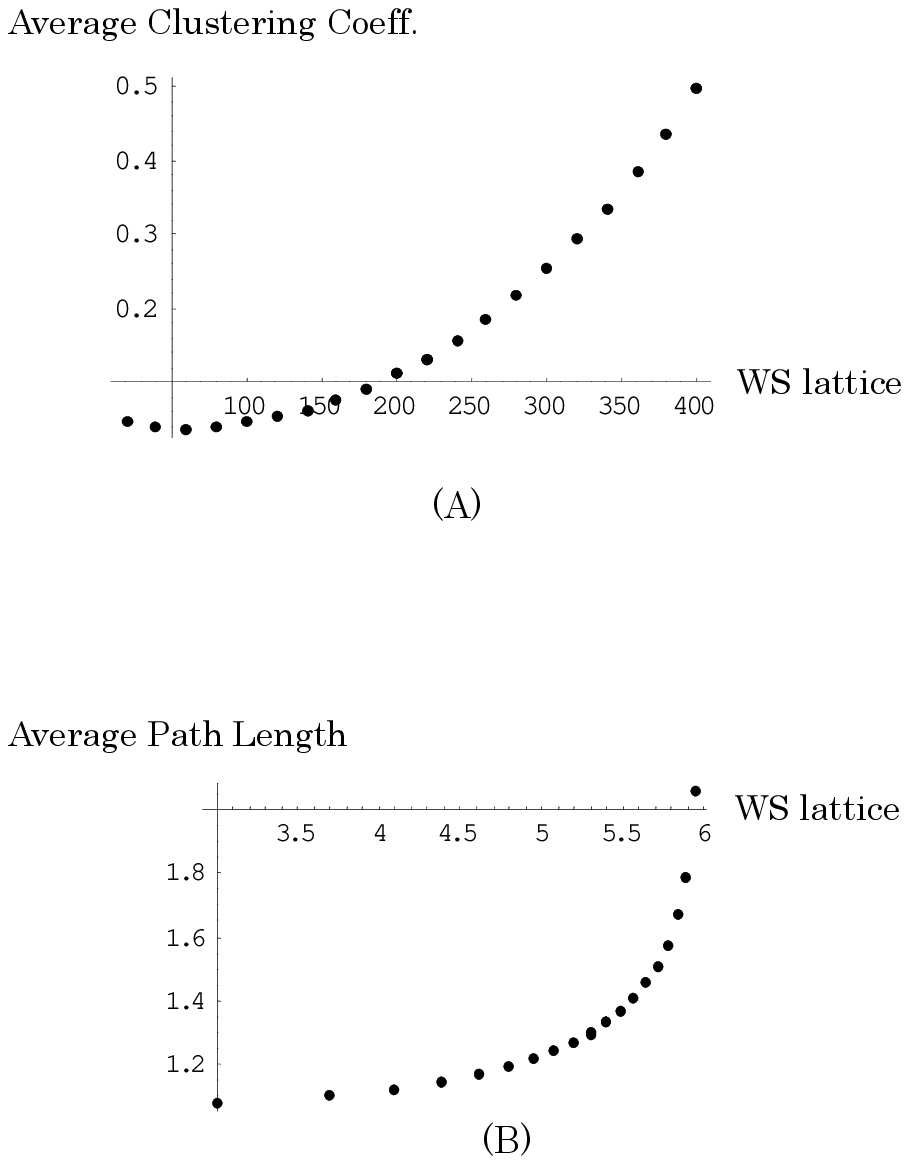} \includegraphics[scale=0.6,clip]{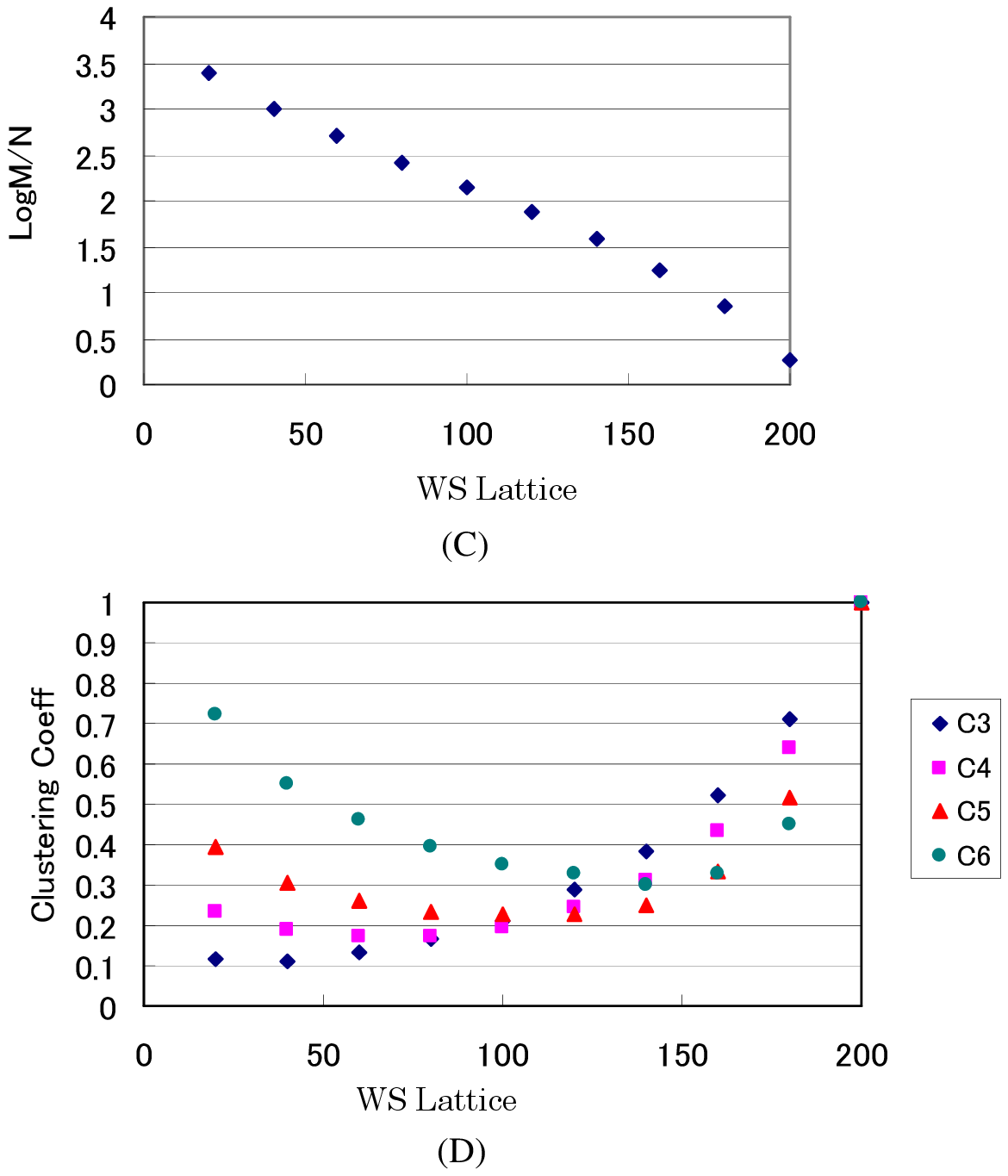} \\
\caption{Average of local clustering coefficient (A),  average path length (B), Milgram condition $M_6/N$ 
and generalized clustering coefficient in the Model (I).  }
\end{center}
%%%%%%%%%%%%%%%%%%%%
\end{figure} 

As the proportion of preferential attachment in the network construction increases, that is, $N-N_0$ to $N_0$, increases, 
the network gets closer to the scale free network.   
Though we find  that $L$ and $C^{(3)}$ increase together from (A) and (B) in fig.6, $M_6/N$ in Milgram condition  
decreases in fig.6 (C), as $N_0$ grows larger. 
In other words, as $C^{(3)}$ increases, $L$ becomes large but $M_6/N$ decreases. 
Thus the increasing of small size closed cycles retard the spreading of information over networks. 
While the same behavior is also found in fig.6 (D), the larger size closed cycles ($C_{(q=5,6)}$) have not any definite influence 
on the spreading of information.    
  $C_{(q)}$ with larger $q$ temporarily decreases and then increases with $N_0$. 
We infer that this behavior of  $C_{(q)}$（$q>3$） makes the increase of $L$ slow in the small $N_0$ area in fig.6 (B). 
Anyway, information spreading over networks is closely bound up with small size cycles. 

These properties are nearly the same as those of small world networks. 
In general small world networks as also shown in fig.1, the increase of closed cycles urges the increase (decrease) of $L$ $(M_n/N$), 
and so closed cycles retards information propagation. 
 
As a consequence, we can not, however, find  any indications of transition phenomena between scale free networks 
and small world networks in this approach,  
where we bring scale free networks (BA model) into small world networks through regular networks (see fig.5).  
When transforming a network from BA model to a regular network (WS lattice), and further from small world  small world to  small world in fig.5, 
$C^{(3)}$ or$C_{(3)}$ become larger and $M_6/N$ gets smaller monotonously between them.

\subsection{Simulation Results of Model II}
\hspace{5mm}
(A) , (B) , (C) and (D) of fig. 7 show  $C^{(3)}$, $L$,  $M_6/N$ and $C_{(q)}$ for $3\leq n \leq 6$ vs. the rewiring ratio $P$ in the model II,  respectively. 
The each plot is estimated in the average over five rounds at $N=500$ for (A) and (B) and the average over five rounds at $N=200$ for (C) and (D) 
by the similar reason to the previous subsection.  
The initial scale free network is made based on BA model with the scaling exponent $\alpha=3$ \cite{Albe2}.  
On the whole, as the rewiring ratio increases,  $M_6/N$ and $C_{(q)}$ increase together in  (C) and (D) of fig.7. 
So $C_{(q)}$,  which conveys the information about how many cycles there are in a network, works to urge information propagation. 
This disposition looks like scale free networks. 
But $C^{(3)}$ becomes larger up to about $P=0.4$ and becomes smaller for larger $P$ than 0.4. 

We analyze $P$ dependent behaviors of the mathematical quantities represented in fig. 7 at full length.    

\begin{enumerate}
\item  $0\leq P \leq 0.2$ \\
 $M_6$, $C_{(q>3)}$ and $L$ are almost constant but $C^{(3)}$ and $C_{(3)}$ grow larger together. 
It is reasonable that $C_{(3)}$ grows larger by the construction.  
So the effect of small size cycles in the networks does not reflect on information propagation.

\item  $0.2 \leq P \leq 0.4$ \\
At a little larger $P=0.2$,   $M_6$ and $C_{(q>3)}$ and $L$ are  leap little bit.   
They are, however,  almost constant basically in this region of $P$. 
The increment of $C^{(3)}$ slows down. 
The value of $L$ shows no great difference  as usual. 

\item  $0.4\leq P \leq 0.6$ \\
In this region of $P$, $M_6$ make a gradual ascent and $L$ make a gradual descent, as $P$ becomes larger.
Though $C_{(q\geq3)}$ are almost constant,  $C^{(3)}$ begins decreasing. 
The relation between $C^{(3)}$ and $M_6$ or $L$ is like that of small world networks. 
$C_{(q\geq3)}$, have not any effect on the behavior of $M_6$ or $L$. 

\item  $0.6 \leq P <1.0$ \\
$M_6$ is on the notable increase and $L$ is on the notable decrease in this region. 
Though $C_{(q\geq3)}$ is increasing,   $C^{(3)}$ is slightly decreasing. 
While the behavior looks like small world  networks from the perspective of $C^{(3)}$, 
the relation looks like scale free networks from the perspective of   $C_{(q\geq4)}$.  
 \end{enumerate}
 
 It is surmised that the reason that the increase of $C_{(3)}$ gradually slows down with $P$ in fig.7 is 
that the triangles constructed so far begin to destroy   
when edges are taken away from a hub at large $P$. 
 $C_{(3)}$ actually increases during small $P$ in fig.7 (D). 
%Can we sat that the network gets closer to a small world network as $P$ becomes larger, 
%though $C_{(q)}$s' decrease with $q$.  

Since increase of  $C_{(q)}$ gives rise to the increase of $M_6$ on the whole, there are a positive correlation between $C_{(q)}$ and $M_6$. 
In the meanwhile, since $C^{(3)}$ increases at $0\geq P \geq 0.4$ with $M_6$ but  contrariwise $C^{(3)}$ deceases at $0.4 \geq P$ with $M_6$.  
However, fig.7 (A) shows a little strange behavior.
Since BA scale free network is brought to "samall world networks" in such meaning 
proposed by Watts-Strogaz in Model II, which have a large clustering coefficient and a small average path length, 
 this fact shows that the average local clustering coefficient $C^{(3)}$ may get some idea of SF-SW transition even at a fixed scaling exponent. 
But note that we can not completely affirm that it is so in the cases, where $C^{(3)}$ and $C_{(3)}$ do not show the same behavior. 
Thus it is considered that $C^{(3)}$ and $C_{(q)}$ have different implications in this phenomena, respectively.  
While the small world network like behaviors in the relation between $L$ and $C^{(3)}$  have been pointed out 
in the tunable scale free model proposed by Holem and Kim\cite{Holme}, the result corresponds to the behavior 
in larger $P$ area in our model.
 
 \begin{figure}[t]
%%%%%%%%%%%%%%%%%%%%%%% 
\begin{center}
\includegraphics[scale=0.85,clip]{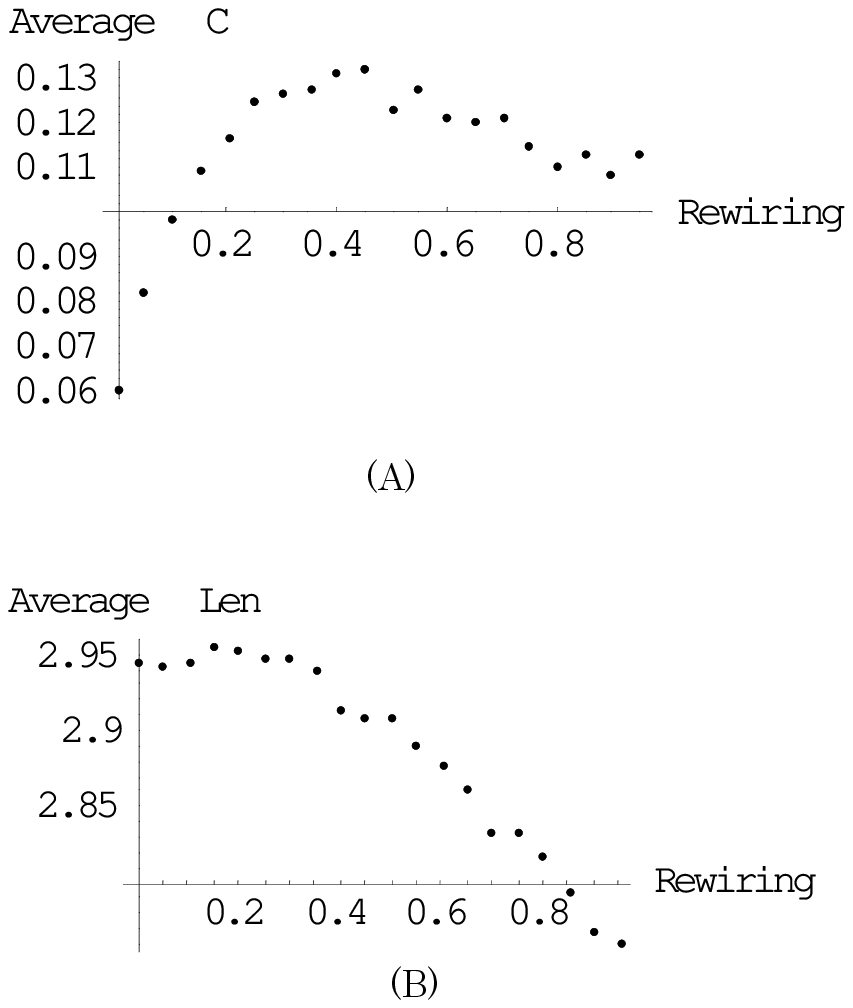} \includegraphics[scale=0.55,clip]{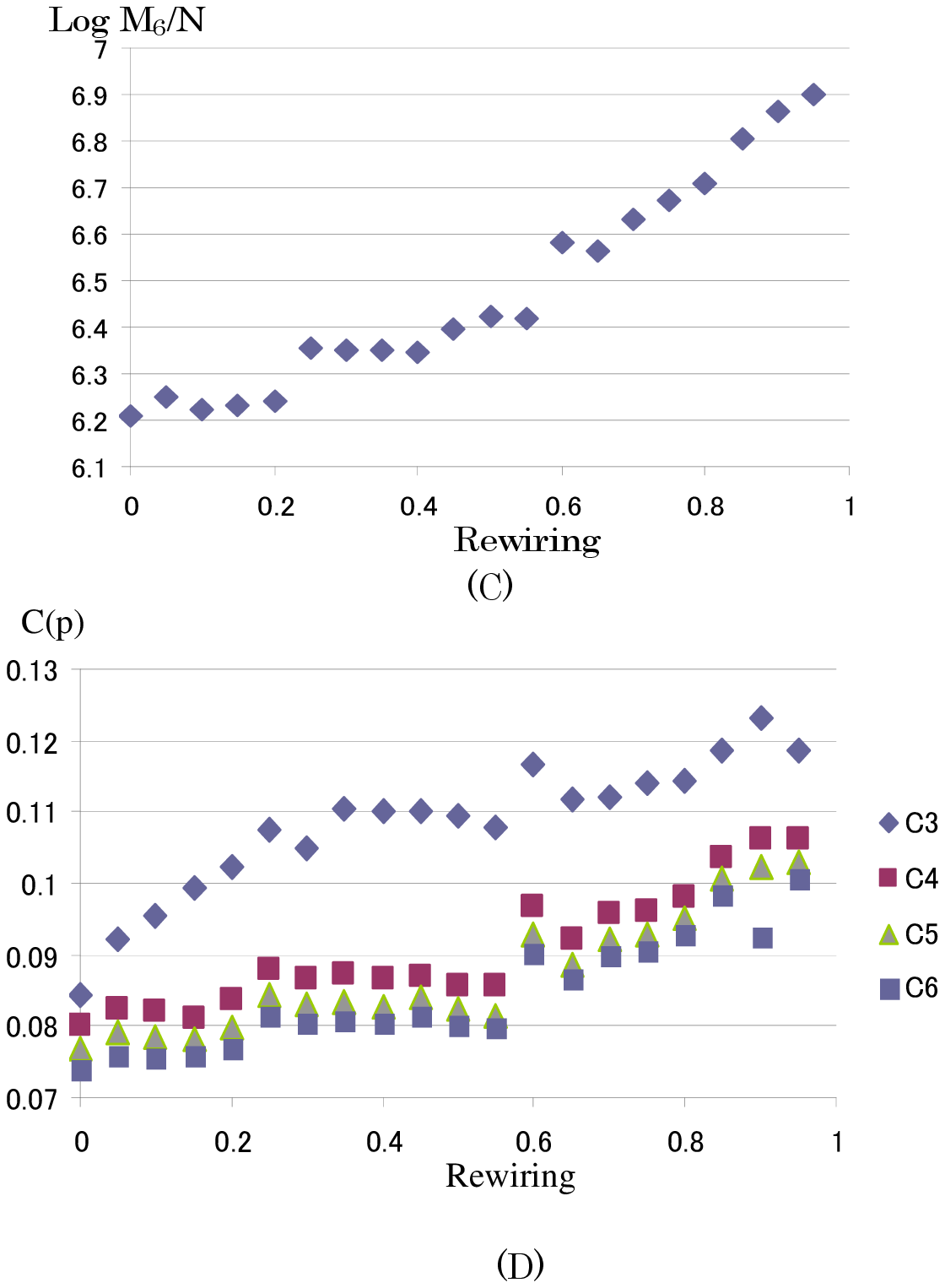} \\
\caption{Average of local clustering coefficient(A),  average path length (B), Milgram condition $M_6/N$ and generalized clustering coefficient in Model (II).  }
\end{center}
%%%%%%%%%%%%%%%%%%%%
\end{figure} 

In Model II,  we can not also see an obvious transition between the scale free network like behaviors and 
the small world network like ones in the effect of closed paths on information propagation.   
Such characteristic of scale free networks observed in fig.2, that is, 
the relation between $L$ and $C_{(q)}$, appeared through varying the scaling exponent of the networks \cite{Toyota10},\cite{Toyota11} 
(note that since the network construction is based on BA model in the Model II (and also I),  
the scaling exponent is the fixed value $3$, even though the networks get near to a scale free network).  
So we speculate that  a crucial point of explaining the fact that the closed paths (larger $C_{(q)}$) 
expedited the propagation of information in scale free networks is this difference in $\alpha$.      
Such characteristic of scale free networks observed in fig.2 could not appear unless the scaling exponent does not change.

Thus we recognize that the average of local clustering coefficient $C^{(3)}$ and transitivity or generalized clustering coefficient $C_{(3)}$ 
are different meanings in networks at least. 
It is an example that the networks considered in this subsection can uncover these differences.   
We further pursuit two networks introduced in this section from the perspective of disorder in the sequent subsection.

%Thus we can not realize the characteristic of scale free networks in these models. 
%%%%%%%%%%%%%%%%%%%%%%%%%%%%%%%%%%%%%%%%%%%%%%%%%%%%%%%%%%%%%%%%
% \begin{figure}[t]
%%%%%%%%%%%%%%%%%%%%%%% 
%\begin{center}
%\includegraphics[scale=0.8,clip]{Fig4AB.eps} \\
%\caption{Clustering Coefficient and average path length in the Model (II).  }
%\end{center}
%%%%%%%%%%%%%%%%%%%%
%\end{figure} 
\subsection{Disorder}
\hspace{5mm}
We estimate the standard deviation and the entropy of degree distribution of two types of networks introduced in previous subsections. 

\underline{Model I}\\
Fig.8 shows that the entropy and the standard deviation decrease as the size $N_0$ of WS lattice becomes larger. 
Then the clustering coefficients $C_{(3)}$ and $C^{(3)}$ enlarged  together as shown in fig.6 (A) and (D). 
Thus the entropy and the standard deviation decrease together when the clustering coefficients become larger. 
Then $L$ increases and $M_6/N$ in Milgram condition decreases.  
Overall properties in the Model I are like those of small world networks as also mentioned in the subsection 3.2. 

\underline{Model II}\\
As shown in fig.9, the standard deviation and the entropy increase as $P$ becomes larger, 
when $C_{(q)}$ becomes on the whole lager in fig.7 (D). 
Then $L$ decreases and $M_6/N$ in Milgram condition increases.  
 These properties are like those of scale free networks. 
It is, however, $C^{(3)}$ had a strange behavior as compared with other indices.  \\ 

So far we have analyzed four types of networks, scale free networks, small world networks, Model I and Model II.   
The standard deviation and the entropy that are considered to measure a sort of disorder increase with $M_6$ in all of them. 
This fact implies that disorder is a crucial index for information propagation in networks.  
     
% Then the generalized clustering coefficients  $C_{(q)}$, however, decrease in fig.7 (D).   
 % Such property is like scale free networks where the standard deviation takes small value at small $C_{(3)}$ in contrast.  
 % As for the entropy,  it shows a little complicated behavior. 
 %Though the entropy $S$ increases when $P$ has small value, S$S$ slowly decreases for large $P$. 
 % What I mean to say is that as  $C_{(q)}$ becomes larger with $P$ at the small $P$ area from fig.7 (D), 
%it looks like small world networks, but  as  $C_{(q)}$ slowly decreases with $P$ at larger $P$ area from fig.7 (D), 
%it looks like scale free networks. 
%On the contrary, we take account of $C^{(3)}$,  $C^{(3)}$ becomes smaller with $P$ at the small $P$ area from fig.7 (A), 
%it looks like scale free world networks, but  as  $C_{(q)}$ increases with $P$ at larger $P$ area from fig.7 (A), 
%it looks like small world networks. 
%The behavior of $S$ indicates the transition between small world like property and scale free like one in either event.  
%So $S$is an important index at transition appeared in the relation to cycle structure-information spreading.   

Notice the construction of Model II only corresponds to the third step in the model introduced by 
B.Wang et al.\cite{Wang}  as mentioned above. 
Due to it, the variation of  $C^{(3)}$ or $C_{(q)}$ is comparatively small. 
By exploring more various clustering tunable scale free networks such as full version of Wang et al.\cite{Wang} 
and others proposed by \cite{Holme},\cite{Zhang}, we should scrupulously examine what is an essential index for the transition. 

 %\begin{figure}[b]
%%%%%%%%%%%%%%%%%%%%%%% 
%\begin{center}
%\includegraphics[scale=0.8,clip]{Milg4.eps} \\
%\caption{   The relation between Milgram condition and $N_0$  in the Model (II).  }
%\end{center}
%%%%%%%%%%%%%%%%%%%%
%\end{figure} 

\begin{figure}[t]
%%%%%%%%%%%%%%%%%%%%%%% 
\begin{center}
\includegraphics[scale=0.8,clip]{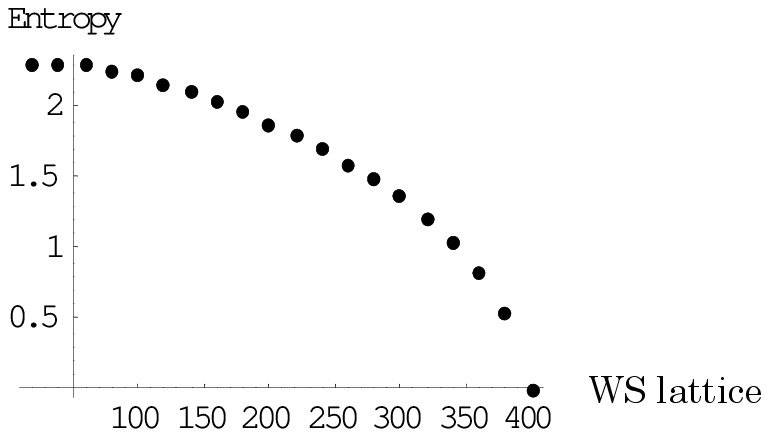}\includegraphics[scale=0.75,clip]{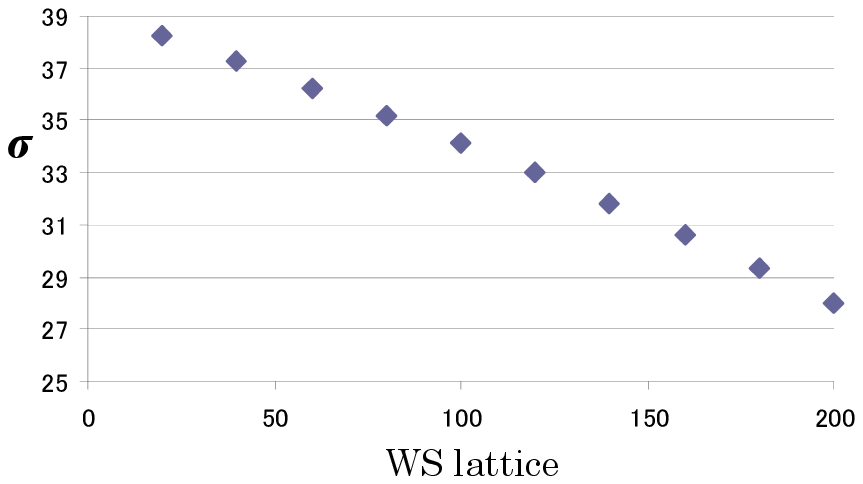}  \\
\caption{Entropy and standard deviation in Model (I).  }
\end{center}
%%%%%%%%%%%%%%%%%%%%
\end{figure} 

\begin{figure}[t]
%%%%%%%%%%%%%%%%%%%%%%% 
\begin{center}
\includegraphics[scale=0.95,clip]{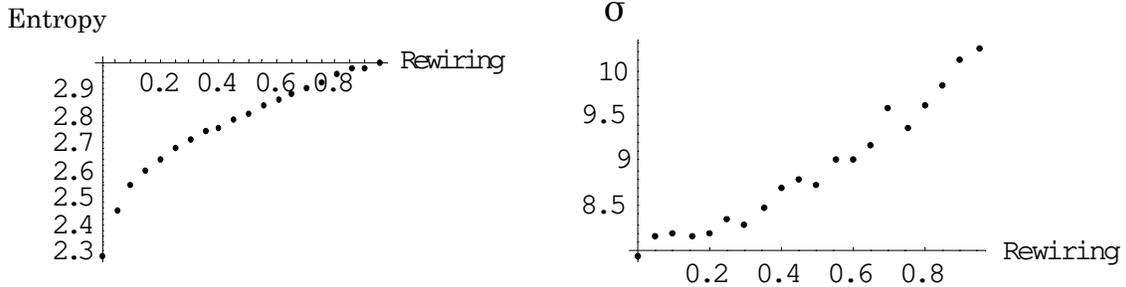} \\
\caption{Entropy and Standard deviation in Model (II).  }
\end{center}
%%%%%%%%%%%%%%%%%%%%
\end{figure}

% \begin{figure}[t]
%%%%%%%%%%%%%%%%%%%%%%% 
%\begin{center}
%\includegraphics[scale=0.7,clip]{Fig6GC.eps} \\
%\caption{Generalized Clustering Coefficient in Model (II).  }
%\end{center}
%%%%%%%%%%%%%%%%%%%%
%\end{figure} 

\section{SUMMARY }
\hspace{5mm}
Milgram condition plays an important role  in the analysis of six degrees of separation\cite{Aoyama}.
We have earlier pointed out that there is a  huge difference in the relation between Milgram condition 
and the generalized clustering coefficients in scale free networks and small world networks \cite{Toyota6},\cite{Toyota7}. 
So we found that cycle structures immanent in a network do not always play  effective role for information propagation.    

In this article, we first investigated the difference and further pursued what is a crucial mathematical quantity.  
As a result of investigating both networks, we find that a sort of "disorder", which is measured by the standard deviation of degree data 
and the entropy of degree distribution, plays more important role for information propagation rather 
than the generalized clustering coefficient  $C_{(q)}$ and the average of local clustering coefficients $C_{(3)}$.   

In this article, we inquired into it in detail based on further two types of intermediate networks. 
One is the network that link a regular network and the scale free networks introduced by Barabashi and Albert\cite{Albe2}. 
Second is  the network that link a sort of small world networks and the scale free network introduced 
by Barabashi and Albertt\cite{Albe2}. 
It would be expected that the transition between  scale free network like property and small world network like one 
may be clearly uncovered by the study of the proposed networks. 
But we could not explicitly understand it.  
There is a positive or negative correlation between $C_{(q)}$ and $M_6$, that is, information propagation on the whole,  
but $C^{(3)}$ has displayed a little strange behavior as compared with  the behaviors of $C_{(q)}$, $L$, $M_6$ and  $C_{(q)}$. 
So $C^{(3)}$ may grasp a part of the transition between scale free $\leftrightarrow$ small world (SF-SW transition).  
$C^{(3)}$ and $C_{(q)}$ have some different functions and important meanings, respectively, 
through studying such types of networks.  
The indeces  $C^{(3)}$ and $C_{(q)}$ would complement each other in our discussions apparently. 

In scale free networks, it was also pointed out that the scaling exponential is crucial for information propagation.  
Furthermore, we found that more essential concept for information propagation is "disorder" 
through studying various type networks as a consequence. 
 Then we found that small entropy of degree distribution and small standard deviation of degree data impede the propagation 
of information all in small world networks, scale free networks and the two types of networks proposed in this article. 

So the behaviors of $M_6/N$ or six degrees of separation in all networks discussed in this article are understood in a unified way by disorder.  
As a result, it turned out that the more essential quantity for information propagation is not the number of closed cycles included 
in a network in general  but the entropy of degree distribution or standard deviation of degree data in the  network.  
We should investigate more wide class of networks to verify our discussions. 

%%%%%%%%%%%%%%%%%%%%%%%%%%%%%%%%%%%%%%%%%%%%%%%%%%%%
 %%%%%%%%%%%%%%%%%%%%%%%%%%%%%%%%%%%%%%%%%%%%%%%%%%%%
 %xxxxxxxxxxxxxxxxxxxxxxxxxxxxxxxxxxxxxxxxxxxxxxxxxxxxxxxxxxx
%%%%%%%%%%%%%%%%%%%%%%%%%%%%%%%%%%%%%%%%%%%%%%%%%%%%
 %%%%%%%%%%%%%%%%%%%%%%%%%%%%%%%%%%%%%%%%%%%%%%%%%%%%

\appendix 
\section{Basis of Power Low Distribution}
\hspace{5mm}
We describe some basic relations which are used in this article in networks with the power low distribution of degree in this appendix.   
\subsection{Normalization}
\hspace{5mm}
For scale free networks, the degree distribution with the normalization constant $c$ is given by $P(k)=ck^{-\alpha}$. 
The normalization factor is determined by the following condition;
\begin{equation}
\sum_{k=k_{min}}^{k_{max}} cP(k)= \sum_{k=k_{min}}^{k_{max}} ck^{-\alpha} =1,
\end{equation}
where the minimal degree $k_{min} =0$ and  the maximal degree $k_{max} =\infty$ are taken in an ideal case but this distribution diverges at $k=0$. 
Thus there must be a lower bound for $k_{min}$.  
From the lower bound $k_{min}$ and eq. (14),  we obtain 
\begin{equation}
P(k)= \frac{k^{-\alpha}}{\zeta (\alpha, k_{min})},
\end{equation}
where 
\begin{equation}
\zeta (\alpha, k_{min}) \equiv \sum_{n=0}^\infty (n+k_{min})^{-\alpha} 
\end{equation}
is the Hurwitz zeta function that is a generalization of the original Riemann zeta function\cite
{Clau}. 
It is, however, difficult to obtain a concrete and general expression for the Hurwitz zeta function.  

The calculation, however, becomes rather simpler when a continuous approximation is made. 
Then the normalization constant and the entropy of degree distribution can be analytically calculated for scale free networks. 
First of all, we determine normalization factor of the degree distribution;
 \begin{equation}
\int_{k=k_{min}}^{\infty} P(k)  dk=\int_{k=k_{min}}^{\infty}  ck^{-\alpha} dk =1. 
\end{equation}
From eq.(17),  we obtain 

\begin{equation}
P(k)= (\alpha -1) k_{min}^{\alpha-1}  \;\; k^{-\alpha}  . 
\end{equation}
Then the average degree is given by
\begin{equation}
<k>=\int^{\infty}_{k=k_{min}} kP(k) dk= \frac{(\alpha -1) k_{min}^{\alpha-1} }{2-\alpha} [k^{2-\alpha}]^{\infty}_{k=k_{min}}.
\end{equation}
For $\alpha>2$, we obtain 
\begin{equation}
<k>= \frac{(\alpha -1) k_{min} }{\alpha-2}.
\end{equation}
 As to $<k^2>$, we obtain
\begin{equation}
<k^2>=\int^{\infty}_{k=k_{min}} k^2P(k) dk= \frac{(\alpha -1) k_{min}^{\alpha-1} }{3-\alpha} [k^{3-\alpha}]^{\infty}_{k=k_{min}}.
\end{equation}
For $\alpha>3$, we obtain 
\begin{equation}
<k^2>= \frac{(\alpha -1) k_{min}^2 }{\alpha-3}.
\end{equation}

Thus the valiance is obtained by
\begin{equation}
<\sigma^2>= \frac{(\alpha -1) k_{min}^2 }{(\alpha-3)(\alpha-2)^2}, 
\end{equation}
which has a meaning only for $\alpha>3$. 

\subsection{Cut off Theory}
\hspace{5mm}
We observed that some moments of degree  $k$ are divergent at small $\alpha$ in previous subsection. 
Since the network size, however,  is finite in real networks,  the maximum of degree would not be $\infty$ but be
limited to be some value. 
Then the normalization factor $c$ should be reevaluated. 
 In continuous limit, this condition is
\begin{equation}
\int_{k_{min}}^{k_{max}} ck^{-\alpha} dk= 1. 
\end{equation}
As a result, we obtain 
\begin{equation}
c= \frac{\alpha-1}{ k_{min}^{1-\alpha} -k_{max}^{1-\alpha} }.
\end{equation}

We estimate the value $k_{max}$.  
A basic idea to do so is the relation;
\begin{equation}
\sum_{k_{max}}^{\infty} p(k) N \sim 1 
\end{equation}
which means that the number of the nodes that have degree no less than $k_{max} $ is one, roughly. 
In continuous limit, this condition leads to
 \begin{equation}
\sum_{k_{max}}^{\infty} p(k) N \sim \int^{\infty}_{k_{max}} p(k) Ndk =N\frac{ k_{max}^{1-\alpha }}{ k_{min}^{1-\alpha}-k_{max}^{1-\alpha  } }\sim 1. 
\end{equation}
By noticing $k_{max} >>k_{min}$,  we obtain the relation 
 \begin{equation}
k_{max} \sim k_{min} N^{\frac{1}{\alpha -1}}. 
\end{equation}

Using this relation, the normalization factor $c$ is obtained by 
\begin{equation}
c= \frac{\alpha-1}{ k_{min}^{1-\alpha} -k_{max}^{1-\alpha} }\sim \frac{\alpha-1}{ k_{min}^{1-\alpha} (1-1/N) }\sim 
\frac{\alpha-1}{ k_{min}^{1-\alpha }}, \;\;\;\mbox{ for large} \;\;N.
\end{equation}

 Using $k_{max}$,
 we reevaluate $<k>$ and $<k^2>$; 
%%%% \begin{align}%\displaystyle 
 \begin{eqnarray}
 <k>&= \int_{k_{min}}^{k_{max}} p(k) k dk= c\frac{ k_{max}^{2-\alpha} -  k_{min}^{2-\alpha} }{2-\alpha}, \\
     && \nonumber\\
 <k^2>&= \int_{k_{min}}^{k_{max}} p(k) k^2 dk= c\frac{ k_{max}^{3-\alpha} -  k_{min}^{3-\alpha} }{3-\alpha}. 
%%%%\end{align}
\end{eqnarray}
Substituting eq.(28) and (29) into eq. (30) and (31),
we obtain
 \begin{eqnarray}
 <k> &\sim& \frac{(\alpha-1)}{2-\alpha} k_{min} (N^{ \frac{ 2-\alpha}{ \alpha-1 } }-1), \\
 <k^2> &\sim& \frac{ \alpha-1 }{  (3-\alpha) } k_{min}^{2}  (N^{ \frac{ 3-\alpha}{\alpha-1} } -1).
\end{eqnarray}
We now consider the cases with $\alpha>1$, and notice 

 \begin{eqnarray}
\frac{2-\alpha}{\alpha-1} &>& 1 \;\;\mbox{for }\; 1<\alpha<2, \\
\frac{3-\alpha}{\alpha-1} &>& 1 \;\;\mbox{for }\; 1<\alpha<3. \\
\end{eqnarray}

%\section*{Acknowledgments}

%\bibliographystyle{ieicetr}
%\bibliography{myrefs}

\end{document}